\documentclass[aps,pra,preprint,groupedaddress,showpacs,%
tightenlines,
nofootinbib,floatfix]{revtex4}

\usepackage{amsmath,amssymb,amsfonts,bm}
\usepackage{graphicx}
\usepackage{dcolumn}
\usepackage{mathrsfs}
\usepackage{hyperref}

\renewcommand{\vec}[1]{\bm{\mathrm{#1}}} 

\begin{document}

\title{Shift and broadening of resonance lines of antiprotonic
  helium atoms in liquid~$^4$He}

\author{Andrzej Adamczak}
\email{andrzej.adamczak@ifj.edu.pl}
\affiliation{Institute of Nuclear Physics, Polish Academy of Sciences,
Radzikowskiego 152, PL-31342~Krak\'ow, Poland}

\author{Dimitar Bakalov}
\email{dbakalov@inrne.bas.bg}
\affiliation{Institute for Nuclear Research and Nuclear Energy,
  Bulgarian Academy of Sciences, Tsarigradsko chauss\'ee 72,
  Sofia 1784, Bulgaria}

\date{\today}

\begin{abstract}
  The shift and broadening of the resonance lines in the spectrum of
  antiprotonic helium atoms $\bar{p}\mathrm{He}^{+}$ located in fluid
  and superfluid $^4$He have been estimated. The contributions to the
  shift and broadening from collective degrees of freedom in liquid
  $^4$He have been evaluated using the phenomenological response
  function. The shift due to collisions of $\bar{p}\mathrm{He}^{+}$ with
  $^4$He atoms has been calculated in the quasistatic limit using the
  experimental pair-correlation function. It has been shown that an
  implanted $\bar{p}\mathrm{He}^{+}$ atom establishes a~good probe of
  liquid-helium properties, since this atom practically does not change
  the target structure.
\end{abstract}


\maketitle

\section{Introduction}
\label{sec:intro}

The exotic antiprotonic helium atoms $\bar{p}\mathrm{He}^{+}$ are formed
when negatively charged antiprotons are slowed down in helium and later
captured by the Coulomb field of the helium nuclei. A fraction of about
3\% of the antiprotons are captured in metastable states with lifetimes
on the order of microseconds, and this has allowed for a series of
high-precision spectroscopy experiments~\cite{yamaz02} that have
produced, among other things, top accuracy values of fundamental
particle characteristics such as the electron-to-antiproton mass ratio
\cite{hori11} and the antiproton dipole magnetic moment~\cite{pask09}.
These high-accuracy goals required various systematic effects to be
accounted for, the density broadening and shift of the spectral lines
being among the most important among them~\cite{tori99}.  The density
effects have been evaluated for a gaseous target in the semiclassical
approach \cite{baka00} using a pairwise interaction potential of an
antiprotonic and an ordinary helium atom, calculated {\em ab initio} in
the frame of the symmetrized Rayleigh-Schr\"{o}dinger theory
\cite{jezi98}, and the results were in agreement with the experimental
data taken at helium gas densities up to 127~g/l~\cite{yamaz02}. The
attempts of the ASACUSA collaboration at CERN for the laser spectroscopy
of antiprotonic atoms in liquid helium~\cite{hori-private} made us
revisit the subject, since the collective degrees of freedom in the
liquid phase (sound waves in the liquid consisting of the neutral $^4$He
atoms) provide a~new mechanism for shifting and broadening the atomic
spectra, in addition to those investigated in gaseous helium. We
expected, however, that collisions of the neutral
$\bar{p}\mathrm{He}^{+}$ atom with the neighboring neutral $^4$He atoms
give the dominant contribution to the shift and broadening, both in the
gaseous and liquid helium targets. The reason was that the
characteristic changes of the $\bar{p}\mathrm{He}^{+}$ energy levels due
to these collisions are a~few orders of magnitude greater than the
typical energy of the sound phonons associated with the momentum
transfer between the laser photons and the liquid.  In the mean time, we
neglected the effects of inelastic collisions of
$\bar{p}\mathrm{He}^{+}$ with $^4$He atoms since the typical thermal
collision energies $\varepsilon_T$ of the order of
$10^{-4}$--$10^{-3}$~eV are much smaller than the separation
$|\Delta\varepsilon|\gtrsim{}0.1$~eV between the energy levels of states
with adjacent values of the quantum numbers. The detailed calculations
of the collisional quenching rate of antiprotonic helium
in~\cite{quench} show that collisional quenching is indeed strongly
suppressed, in agreement with experimental data.

We focus on the transitions
$|i\rangle=(n,\ell)=(39,35)\to|f\rangle=(n',\ell')=(38,34)$
(transition~I) and $(37,34)\to(36,33)$ (transition~II), which are of
major interest for the experimentalists~\cite{hori11}. For transition~I,
the corresponding photon wavelength equals
$\lambda_0=5972.570$~\AA~\cite{tori99} and the resonance energy is
$E_0=2.07589$~eV. The natural width~$\Gamma_n$ is determined by the
Auger decay rate $R_A$ of the final state
\begin{equation}
  \label{eq:gam_Auger}
  \Gamma_n = \hbar/\tau_n \approx \hbar R_A  \,.
\end{equation}
For this line, the experimental rate is
$R_A\approx{}1.11\times{}10^8\mathrm{s}^{-1}$~\cite{yamag02} and
therefore $\Gamma_n\approx{}0.73\times{}10^{-7}$~eV, the corresponding
frequency is $\nu_n\approx{}0.018$~GHz, and the lifetime
$\tau_n=9.0$~ns. In the case of transition~II, the analogous parameters
are: $\lambda_0=4707.220$~\AA~\cite{tori99}, $E_0=2.63392$~eV,
$R_A\approx{}2.2\times{}10^8\mathrm{s}^{-1}$~\cite{hori98},
$\Gamma_n\approx{}1.4\times{}10^{-7}$~eV, $\nu_n\approx{}0.035$~GHz, and
$\tau_n=4.5$~ns.

In Sec.~\ref{sec:liquid_he} we report the estimations of the line
shift and broadening due to the collective motion in liquid
$^4$He, using the Van Howe formalism~\cite{vanh54}. In
Sec.~\ref{sec:single-atom}, we evaluate the line shift due to
collisions of $\bar{p}\mathrm{He}^{+}$ with $^4$He atoms by making
use of the quasistatic limit of the results of Ref.~\cite{baka00}
in a form that allows for exploiting the experimental data on the
pair correlation function in liquid~$^4$He. Unfortunately, this
method cannot be extended to the evaluation of the line
broadening. Section~\ref{sec:concl} includes a brief discussion of
the results.

\section{Line shift and broadening due to the collective dynamics of
liquid helium}
\label{sec:liquid_he}

In this section, the contributions the line shift and broadening due to
the dynamics of liquid $^4$He are estimated in terms of the
quantum-mechanical response function~$\mathcal{S}(\vec{q},\omega)$,
which was introduced by Van Hove~\cite{vanh54} and discussed in detail
for various targets in many textbooks (see e.g., Ref.~\cite{love84}).
The quantities $\hbar\omega$ and $\hbar\vec{q}$ denote, respectively,
the energy and momentum transfers to a~given target. The response
function depends on the target properties at a~fixed temperature and
density. On the other hand, $\mathcal{S}(\vec{q},\omega)$ is independent
of the nature of interaction of an impinging particle with the target.

The influence of liquid $^4$He dynamics on the line shift and
broadening can be estimated using the method developed by Singwi
and Sj\"olander~\cite{sing60} for describing the $\gamma$ quantum
absorption or emission by a~nucleus located in a~condensed target.
According to Ref.~\cite{sing60}, the cross section~$\sigma$ for
photon absorption or emission can be expressed in the simplified
form
\begin{equation}
  \label{eq:sig_abs}
  \sigma(E) = \frac{\mathcal{A}}{\hbar}\, \mathcal{S}_i(q,\omega) \,,
\end{equation}
when the natural resonance width~$\Gamma_n$ is so small that the
analogous cross section $\sigma_0$ for the nucleus set at a~fixed
position can be approximated by the $\delta$-function profile
\begin{equation}
  \label{eq:delta_profile}
  \sigma_0(E) = \mathcal{A}\, \delta(E-E'_0) \,,
\end{equation}
in which $\mathcal{A}$ represents the strength of the resonance, $E$ is
the energy of an absorbed or emitted photon, and $E'_0$ is the resonance
energy. The function $\mathcal{S}_i(q,\omega)$ denotes the Van Hove
single-particle response function, which is the incoherent fraction of
the total response function~$\mathcal{S}(q,\omega)$. In the case of
laser-stimulated transitions in the antiprotonic helium
\begin{equation}
  \label{eq:transf}
  \hbar q = p \,, \qquad \hbar\omega  = E-E'_0 \,,
\end{equation}
where $p$ is the absolute value of momentum of the absorbed or emitted
photon. The resonance energy $E'_0=E_0+\Delta{}E_0$ includes here the
line shift $\Delta{}E_0$ due to the pairwise interaction. Since
$\Delta{}E_0\ll{}E_0$ in liquid helium, as shown in
Sec.~\ref{sec:single-atom}, one can take $E'_0\approx{}E_0$ in numerical
estimates of the collective effects.

In the case of a perfect gas, a harmonic solid, or a particle diffusing
in a classical fluid according to the Langevin equation, the response
function can be rigorously derived~\cite{vanh54}. However, an exact form
of the response function in normal fluid (He~I) and superfluid helium
(He~II) is not known yet.  Therefore, $\mathcal{S}(q,\omega)$ for liquid
helium is usually expressed in terms of simple analytical functions,
such as the Gaussian or Lorentzian function, with several parameters to
be determined in experiments. Extensive measurements of the response
function for liquid helium at various conditions were performed for many
years, using x-ray and neutron scattering. The experimental data are
available only for the momentum transfers $q\gtrsim{}0.1$~\AA$^{-1}$. In
the case of the $\bar{p}^4\mathrm{He}^{+}$ atom, the characteristic
momentum transfer equals $q=p/\hbar=2\pi/\lambda_0$, which gives
$q=0.0011$~\AA$^{-1}$ for transition~I and $q=0.0013$~\AA$^{-1}$ for
transition~II. This is two orders of magnitude smaller than in the case
of neutron-scattering experiments. Nevertheless, some conclusions can be
drawn using the available analytical models and experimental parameters.
It is well known from theory and experiment that the response function
for liquid $^4$He contains contributions from one-phonon and multiphonon
processes.  A~dispersion relation for low-energy phonons in superfluid
$^4$He was first proposed by Landau~\cite{land41a,land41b}
\begin{equation}
  \label{eq:phon_disp}
  \omega_\mathrm{pho} = c_s \, q_\mathrm{pho} \,,
\end{equation}
in which $\omega_\mathrm{pho}$ is the phonon energy, $q_\mathrm{pho}$
denotes the phonon momentum and $c_s$ represents the sound velocity in
the target. Therefore, the interaction of a~photon with a
$\bar{p}^4\mathrm{He}^{+}$ atom located in liquid $^4$He can lead to the
simultaneous creation or annihilation of phonons with the small
energy~$E_\mathrm{pho}$
\begin{equation}
  \label{eq:E_pho}
  E_\mathrm{pho} = \hbar c_s q = c_s p = h c_s /\lambda \,.
\end{equation}
The photon wavelength~$\lambda$ includes here the line shift due to the
pairwise potential.

For many years, a~sharp one-phonon peak observed in the experimental
response function at low temperatures was ascribed to the superfluid
fraction of liquid $^4$He. This fraction is connected with the presence
of the Bose condensate. Therefore, such a~peak was expected to disappear
above the phase-transition temperature $T_\lambda=2.17$~K~\cite{wood78}.
However, the further extensive measurements at
$q\approx{}0.4$~\AA$^{-1}$~\cite{talb88,stir90,bogo04} proved that the
well-defined one-phonon peak does not disappear at $T>T_\lambda$. This
peak was then ascribed to the collective zero-sound mode, which is
independent of the presence of superfluidity. A similar mode is observed
in various classical liquids, at sufficiently low~$q$. Since the laser
spectroscopy of antiprotonic helium is characterized by very low
momentum transfers, the collective-mode contributions to the line shift
and broadening in the He~I region are also determined by this acoustic
one-phonon process.

The low-$q$ phonon processes in liquid~$^4$He are well described by the
following phenomenological one-phonon response
function~\cite{talb88,stir90}
\begin{equation}
  \label{eq:S_1}
  \begin{split}
  \mathcal{S}_1(q,\omega) =  \frac{\hbar}{\pi} [n_\mathrm{B}(\omega,T)+1]
  Z(q,T) \Biggl[ & \frac{\Gamma_1(q,T)}
  {\hbar^2[\omega-\omega_\mathrm{pho}(q,T)]^2+\Gamma_1^2(q,T)} \\
  & -\frac{\Gamma_1(q,T)}
  {\hbar^2[\omega+\omega_\mathrm{pho}(q,T)]^2+\Gamma_1^2(q,T)}
  \Biggr]  \,,
  \end{split}
\end{equation}
in which $n_\mathrm{B}$ is the Bose factor for phonons
\begin{equation}
  \label{eq:Bose_fac}
  n_\mathrm{B}(\omega,T) = [\exp(\beta_T\omega)-1]^{-1} ,
  \qquad  \beta_T = (k_\mathrm{B} T)^{-1}
\end{equation}
and $k_\mathrm{B}$ is the Boltzmann constant. The one-phonon intensity
is denoted here by $Z(q,T)$ and $\Gamma_1(q,T)$ represents the half
width at half maximum of the one-phonon peak. Equation~(\ref{eq:S_1})
includes both the one-phonon creation and annihilation. The experimental
functions $Z(q,T)$, $\omega_\mathrm{pho}(q,T)$, and $\Gamma_1(q,T)$ for
the saturated vapor pressure (SVP) are presented in Ref.~\cite{stir90}
($q=0.4$~\AA$^{-1}$) and~Ref.~\cite{bogo04} ($q=0.22$~\AA$^{-1}$). In
general, for such values of~$q$, these parameters are insensitive to the
phase transition from He~II to~He~I. The phonon energy
$\omega_\mathrm{pho}$ practically does not change with temperature in
superfluid helium and in the vicinity of~$T_\lambda$. The same is true
for the peak intensity $Z(q,T)$. The width~$\Gamma_1$ rises
exponentially with rising temperature below $T\lesssim{}2.3$~K and there
is no abrupt change at~$T_\lambda$. The smallest width reported for
$q=0.4$~\AA$^{-1}$~\cite{stir90} is on the order of~1~GHz at about~1~K.
However, this result is entangled with the experimental energy
resolution of 20~GHz, which is taken into account using a~convolution of
the function~(\ref{eq:S_1}) with the Gaussian that describes the
resolution. The more accurate and reliable measurements~\cite{meze83}
using the neutron spin echo techniques give $\Gamma_1<1$~GHz at 1.35~K,
for $q=0.4$~\AA$^{-1}$. A~similar result is reported in
Ref.~\cite{klim03}. For $T\gtrsim2.3$~K, the width
$\Gamma_1$($q=0.4$~\AA$^{-1}$) increases slowly from 40 to 70~GHz at
$T\approx{}4$~K. Let us note that the observed behavior of $\Gamma_1$ as
a~function of temperature is connected with increasing randomness of the
considered system (see e.g., Ref.~\cite{alex09} and references therein).

In the antiprotonic helium case, the momentum transfers are smaller by
a~factor of 220--400 than the lowest experimental $q$ from
Refs.~\cite{stir90,bogo04,meze83,klim03}. Thus, the knowledge on the
behavior of $\Gamma_1$ as a~function of small $q$ is very important. The
experimental data presented in Ref.~\cite{bogo04} show that $\Gamma_1$
linearly decreases towards very small values with decreasing~$q$, for
$q\lesssim{}0.7$~\AA$^{-1}$. This is consistent with theory which
predicts that the one-phonon peak has the $\delta$-function profile and
$\Gamma_1\to{}0$, in the limit $q\to{}0$. Using the linear
proportionality and the above-mentioned experimental data, one obtains
$\Gamma_1<10^{-2}$~GHz for $T\lesssim{}1$~K. Above this temperature,
$\Gamma_1$ exponentially rises to 0.1~GHz at 2.3~K and then slowly
increases to about 0.18~GHz at~4~K.

Since in the antiprotonic-helium experiments
$\beta_T\omega=\beta_T\omega_\mathrm{pho}\sim{}10^{-2}\ll{}1$, the
phonon population factor $n_\mathrm{B}(\omega,T)+1$
in~Eq.~(\ref{eq:S_1}) approximately equals $1/(\beta_T\omega)$ for the
phonon creation and $-1/(\beta_T\omega)$ for phonon annihilation. As
a~result, assuming that
$\mathcal{S}_i(q,\omega)\approx\mathcal{S}_1(q,\omega)$ is a~reasonable
approximation, the total photon cross section~(\ref{eq:sig_abs}) can be
expressed as follows
\begin{equation}
  \label{eq:sig_abs_1}
  \sigma(E) = \frac{\mathcal{A}}{\pi} \, \frac{Z(q,T)}{\beta_T\omega}
  \Biggl[ \frac{\Gamma_1}
  {[E-(E'_0+E_\mathrm{pho})]^2+\Gamma_1^2} \\
   +\frac{\Gamma_1}
  {[E-(E'_0-E_\mathrm{pho})]^2+\Gamma_1^2}
  \Biggr] .
\end{equation}
Thus, the resonance line is split into the two lines, which are
characterized by the line shifts $\Delta{}E_1=E_\mathrm{pho}$ and
$-E_\mathrm{pho}$.

The absolute values of the line shifts $|\Delta{}E_1|=E_\mathrm{pho}$,
which were calculated using Eq.~(\ref{eq:E_pho}), are shown in
Table~\ref{table:E_pho} as functions of temperature.
\begin{table}[htb]
  \begin{center}
    \caption{The absolute values of line shift $\Delta{}E_1$ in
      superfluid and fluid $^4$He at SVP as functions of temperature.}
    \label{table:E_pho}
    \begin{ruledtabular}
    \newcolumntype{.}{D{.}{.}{2.2}}
    \begin{tabular}{. . .}
      \multicolumn{1}{c}{Temperature}&
      \multicolumn{2}{c}{Line shift [GHz]}\\
      \cline{2-3}
      \multicolumn{1}{c}{[K]}&
      \multicolumn{1}{c}{Line~I}&
      \multicolumn{1}{c}{Line~II}\\
      \hline
      1.20 & 0.40 & 0.50 \\
      1.50 & 0.39 & 0.50 \\
      1.75 & 0.39 & 0.49 \\
      2.00 & 0.39 & 0.48 \\
      2.17 & 0.36 & 0.46 \\
      2.20 & 0.37 & 0.47 \\
      2.50 & 0.37 & 0.47 \\
      3.60 & 0.35 & 0.44 \\
      4.00 & 0.32 & 0.40 \\
      4.22 & 0.30 & 0.38 \\
  \end{tabular}
  \end{ruledtabular}
  \end{center}
\end{table}%
The velocities~$c_s(T)$ at SVP were taken from Ref.~\cite{mayn76} for
He~II and from Ref.~\cite{find38} for He~I. As it was observed in
experiments, the calculated phonon energies do not significantly change
in He~II and in the vicinity of $T_\lambda$. The calculated ratio
$\Gamma_1/|\Delta{}E_1|$ equals about 0.02 at $T\approx{}1$~K and
increases to about 0.5 at~4~K.

\section{Collisional shift of resonance lines}
\label{sec:single-atom}

In Refs.~\cite{baka00,hfi12} the density shift and broadening of the
spectral lines of antiprotonic helium atoms in gaseous helium were
evaluated in the frame of the semiclassical approach of
P.W.~Anderson~\cite{anderson}.  In this approach the emitter --- the
antiprotonic atom --- is subject to full scale quantum treatment, while
the perturber --- the ordinary helium atom --- evolves classically. The
very good agreement of the theoretical results of
Refs.~\cite{baka00,hfi12} with the experimental data taken at a broad
range of helium gas %
%
densities up to 127~g/l
should be attributed to
(1) the use of an accurate pair-wise state-dependent potential for the
interaction of antiprotonic and ordinary helium atoms, calculated
{\em ab initio} with the symmetrized Rayleigh-Schr\"{o}dinger theory
\cite{jezi98};
(2) the use of curvilinear classical trajectories of the
perturbers determined by the interaction potential for the
{\em initial state} of the transition; and
(3) the fact that the conditions which justify the impact approximation
and the approximation of binary collisions --- typical collision
duration smaller by an order of magnitude than the average interval
between collisions, emitter excitation energies much larger than the
thermal collision energies, uncorrelated motion of the perturbers, etc.\
--- are satisfied for the densities and temperatures in consideration.
In liquid helium, however, the target density may be still higher, the
motion of helium atom cannot be considered as uncorrelated, and this
approach cannot be applied. We therefore put the results of
Ref.~\cite{baka00} in a form that allows us to use phenomenological data
about the liquid helium target density instead of the theoretical
calculations, and this way we obtain a reliable estimate of the
collisional shift of the spectral lines.  Unfortunately, the line
broadening cannot be evaluated this way since similar naive approaches
are known to produce wrong values, as pointed out in Ref.~\cite{alex09}.

In the semiclassical approach, the density shift $\Delta{}E_0$ of the
resonance energy $E_0$ of the transition $|i\rangle\rightarrow|f\rangle$
between the initial and final quantum states of the antiprotonic atom,
due to the interaction with the atoms of the surrounding helium gas, is
given by
\begin{equation}
  \Delta E_0=N_0\, \bigg\langle 2\pi v\int \mathrm{d}b\, b \,
  \sin\!\left(\int \mathrm{d}t\, \Delta V(\vec{R}_b(t))\right)
  \bigg\rangle_{\! v} \,.
 \label{prl0}
\end{equation}
Here $N_0$ is the number density of the helium gas,
$N_0=\varrho/M_\mathrm{He}$, $\varrho$ being the target density and
$M_\mathrm{He}$ being the $^4$He-atom mass;
$\Delta{}V(\vec{r})=V_f(\vec{r})-V_i(\vec{r})$ is the difference of the
state-dependent $\bar{p}^4\mathrm{He}^{+}$-He interaction potentials;
$\vec{R}_b(t)$ is the classical trajectory of a~He atom with impact
parameter~$b$, determined by the interaction potential in the %
{\em initial} state and parametrized with the proper time~$t$; and
$\langle\ldots\rangle_v$ denotes averaging over the Maxwell-distributed
asymptotic velocities~$v$ of the helium atom. Equation~(\ref{prl0}) has
been derived in the approximation of an ideal helium gas, pairwise
$\bar{p}^4\mathrm{He}^{+}$-He interaction, and under a~few more
assumptions discussed in detail in Ref.~\cite{baka00}.
At low target temperatures, kinetic energy of most of the incident
helium atoms is small, and also small is the phase accumulated along
a~typical trajectory
\begin{equation}
  \eta_b=\int \mathrm{d}t\,\Delta V(\vec{R}_b(t))\ll 1 \,,
\end{equation}
so that $\sin\eta_b\approx\eta_b$. We can then transform
Eq.~(\ref{prl0}) to the following form
\begin{equation}
  \label{qst}
  \Delta E_0 = \int \mathrm{d}^3 r\,\rho(\vec{r})
  \left[ V_f(\vec{r}) - V_i(\vec{r}) \right] \,,
\end{equation}
where
\begin{equation}
  \label{cdens}
  \rho(\vec{r})=2\pi N_0 \bigg\langle \int \mathrm{d}b \, b
  \int \mathrm{d}t \, \delta(\vec{r}-\vec{R}_b(t)) \bigg\rangle_{\! v}
  \,.
\end{equation}
For spherically symmetric $V_{i,f}(\vec{r})=V_{i,f}(r)$, $\rho$~is also
symmetric: $\rho(\vec{r})=\rho(r)$. Equation~(\ref{qst}) has a simple
physical interpretation: due to the interaction with a helium atom at
position~$\vec{r}$, the energy levels of the initial and final states of
$\bar{p}^4\mathrm{He}^{+}$ (placed at the origin) are shifted by
$V_{i,f}(\vec{r})$, respectively, and the transition energy is shifted
by their difference.  The observable shift is the average of the latter
over the spatial distribution $\rho(\vec{r})$ of the surrounding helium
atoms.  This approximation will be referred to as ``quasistatic limit''.

\begin{figure}[htb]
  \begin{center}
    \includegraphics[width=8cm]{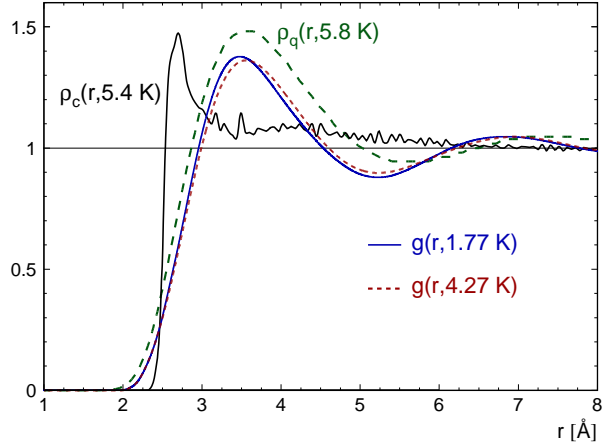}
    \caption{(Color online) The experimental functions~$g(r,T)$ for
      liquid $^4$He at~SVP and $T=$~1.77 and 4.27~K~\cite{sear79},
      together with the classical $\rho_c(r)$ and quantum-mechanical
      $\rho_q(r)$ probability densities which were calculated in the
      two-particle approximation~\cite{baka13}. To emphasize the
      similarity of their shape, the curves $\rho_c(r)$ and $\rho_q(r)$
      in the plot have been {\em renormalized} to unit radial density at
      $r=10$~\AA.}
      \label{fig:dens}
  \end{center}
\end{figure}

Equation~(\ref{qst}) may be the starting point for a~self-consistent
approximate evaluation of the density shift, provided that the helium
gas density $\rho(\vec{r})$ is known. One could think of three different
estimates of $\rho({\mathbf r})$:
\begin{enumerate}
\item
  The helium gas density $\rho_c(r)$ calculated from the classical
  trajectories used in Ref.~\cite{baka00} is one estimate.
  Equation~(\ref{cdens}) gives the algorithm of calculating $\rho_c(r)$
  from the set of classical trajectories.  The corresponding curve for
  a~temperature of 5.4~K, renormalized to~1 at $r=10$~\AA{}, is shown in
  Fig.~\ref{fig:dens}. [The curliness of $\rho_c$ is due to the too
  rough discretization of the integrals in Eq.~(\ref{cdens})].
\item
  Another estimate is the helium gas density $\rho_q(r)$ equal to the
  modulus squared of the two-body scattering wave function for the
  system of point-like helium and $\bar{p}^4\mathrm{He}^{+}$ atoms
  interacting via the potential in the initial state $V_i(r)$. For
  a~comparison, in Fig.~\ref{fig:dens} we plot $\rho_q(r)$
  evaluated~\footnote{The details of the calculation will be presented
    in a separate paper} for the helium temperature $T=5.8$~K and
  renormalized to~1 at $r=10$~\AA~\cite{baka13}.
\item
  A~third estimate is the helium gas density $\rho_\mathrm{exp}(r)$ from
  experiment. Of course, there are no data on the helium density in the
  neighborhood of an $\bar{p}^4\mathrm{He}^{+}$ atom, but as a first
  approximation one can use the static pair correlation function
  $g(\vec{r},T)$ for pure helium that gives the probability density of
  finding a~helium atom located at $\vec{r}$ if the reference particle
  is placed at the origin. At large $r$, $g(\vec{r},T)\to{}1$. Isotropic
  media, such as liquid helium, can be described using the radial
  function $\rho_\mathrm{exp}(r)=g(r,T)$. The pair correlation function
  is usually determined by means of x-ray or neutron scattering.
\end{enumerate}

In the present work we evaluate the collisional line shift using
Eq.~(\ref{qst}) and the phenomenological density $\rho_\mathrm{exp}(r)$
extracted from the set of functions $g(r,T)$ of Ref.~\cite{sear79} which
were determined for fluid and superfluid $^4$He at the saturated vapor
pressure and various temperatures. We make no use of $\rho_c(r)$ or
$\rho_q(r)$ since they have been calculated under assumptions that may
not be valid in liquid helium; any partial results obtained with
$\rho_c(r)$ or $\rho_q(r)$ are listed uniquely for comparison. The
functions $g(r,T)$ for temperatures $T=1.77$ and 4.27~K are plotted in
Fig.~\ref{fig:dens}.
This figure shows that the density distributions $\rho_\mathrm{exp}(r)$
in both the fluid and superfluid~$^4$He are very close. The helium atoms
cannot be closer than about~2~\AA{}. The maxima at about 3.5 and
6.8~\AA{} correspond to the first and second shell of neighbors,
respectively. The shape of the quantum-mechanical density $\rho_q(r)$,
plotted for comparison, is very similar, in particular at small
distances $r\lesssim{}2.5$~\AA. This proves that radial pair correlation
function, which was obtained for a pure $^4$He target, can be applied
for calculating the line shift with the help of Eq.~(\ref{qst}). At
distances $r\gtrsim{}2.5$~\AA, the difference between the functions
$g(r,T)$ and $\rho_q(r)$ increases. This is due to the increasing role
of many-atom interactions which are not taken into account in
$\rho_q(r)$, while the experimental $\rho_\mathrm{exp}(r)$ describes the
real structure of the quantum liquid~$^4$He.

\begin{figure}[htb]
  \centering
  \includegraphics[width=8cm]{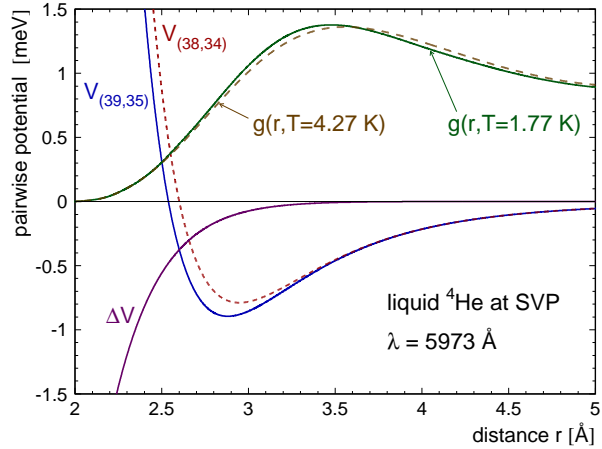}
  \caption{(Color online) The pairwise potential curves
    $V_{(n,\ell)}(r)$, $V_{(n',\ell')}(r)$, and $\Delta{}V(r)$ for
    transition~I, together with the pair correlation function~$g(r,T)$
    at 1.77~K and 4.27~K~\cite{sear79}, versus the distance~$r$ between
    the antiprotonic helium and the $^4$He atom.
  \label{fig:gpot}}
\end{figure}

The potential curves $V_{(n,\ell)}(r)$, $V_{(n',\ell')}(r)$, and
$\Delta{}V(r)$, together with the correlation functions~\cite{sear79},
are shown in~Fig.~\ref{fig:gpot} for transition~I.
One can see that the main contribution to the resonance energy
shift~Eq.~(\ref{qst}) comes from the interval
2.2~$\gtrsim{}r{}\lesssim$~3.4~\AA, where both the $\Delta{}V(r)$ and
$\rho_\mathrm{exp}(r)\equiv{}g(r,T)$ have significant magnitudes.

In order to check the validity of using the pairwise interaction
potentials for the determination of resonance shifts,
\begin{figure}[htb]
  \centering
  \includegraphics[width=8cm]{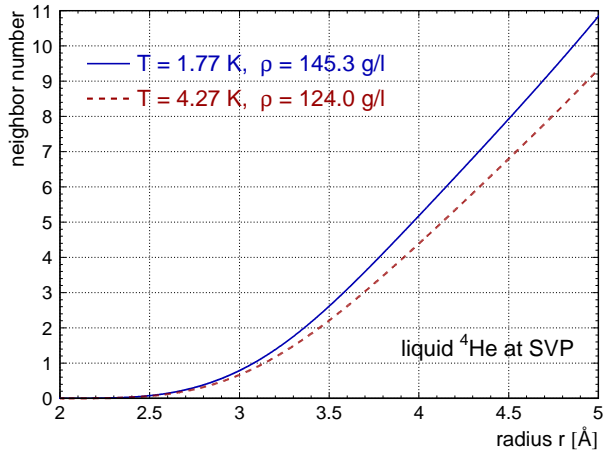}
  \caption{(Color online) The number $n(r)$ of $^4$He atoms within the
    sphere of radius~$r$ that surround an atom placed at $r=0$ in liquid
    helium.}
  \label{fig:neighbors_sea}
\end{figure}%
in~Fig.~\ref{fig:neighbors_sea} we plot the average number $n(r)$ of
$^4$He atoms that are located within the sphere of radius~$r$
\begin{equation}
  \label{eq:neighbors}
  n(r) = 4\pi N_0 \int_0^r \mathrm{d}r'\, r'^2 \rho_\mathrm{exp}(r')\,,
\end{equation}
for $T=1.77$ and 4.27~K at~SVP. The plot shows that $n(r)\leq{}1$ for
$r\leq{}3.1$--3.2~\AA, depending on the target density. Since
$\Delta{}V(r)$ has the largest absolute values within this interval
of~$r$, one can expect that Eq.~(\ref{qst}) establishes a~good
approximation to the resonance shifts.

The number density~$N_0$ in~Eq.~(\ref{cdens}) is a~factor that suggests
a~direct proportionality of the energy shift with respect to the helium
density.  However, the function $\rho_\mathrm{exp}(r)=g(r,T)$ slightly
varies with temperature and thus also with the density, which at SVP is
determined by temperature.  Therefore, the integration
in~Eq.~(\ref{qst}) leads to a higher-order correction to the
above-mentioned linear proportionality.

The resonance line shifts, which were calculated in the quasistatic
limit using the experimental helium density
$\rho_\mathrm{exp}(r)\equiv{}g(r,T)$ from~Ref.~\cite{sear79}, are shown
in Table~\ref{table:shift_sea}
\begin{table*}[htb]
  \begin{center}
    \caption{The line shift $\Delta{}E_0$ [GHz] and the reduced line
      shift $\Delta{}E_0/\varrho$ [GHz.l/g]) for liquid $^4$He at~SVP,
      calculated with Eq.~(\ref{qst}) using $g(r,T)$
      from~Ref.~\cite{sear79}. For comparison, in the last two columns
      are given the values of the reduced line shift in gaseous helium,
      calculated in the semiclassical approach of Ref.~\cite{baka00}.}
    \label{table:shift_sea}
    \begin{ruledtabular}
    \newcolumntype{.}{D{.}{.}{3.3}}
    \begin{tabular}{. . . . . . . .}
      \multicolumn{1}{c}{Temperature}&
      \multicolumn{1}{c}{Density}&
      \multicolumn{2}{c}{$\Delta{}E_0$}&
      \multicolumn{2}{c}{$\Delta{}E_0/\varrho$}&
      \multicolumn{2}{c}{$\Delta{}E_0/\varrho$ in $^4$He gas}\\
      \cline{3-4} \cline{5-6} \cline{7-8}
      \multicolumn{1}{c}{[K]}&
      \multicolumn{1}{c}{[g/l]~~~}&
      \multicolumn{1}{c}{Line~I}&
      \multicolumn{1}{c}{Line~II}&
      \multicolumn{1}{c}{Line~I}&
      \multicolumn{1}{c}{Line~II}&
      \multicolumn{1}{c}{Line~I}&
      \multicolumn{1}{c}{Line~II}\\
      \hline
      1.00 & 145.1 & -64.1 & -37.7 & -0.442 & -0.260 & -0.509 & -0.181 \\
      1.38 & 145.1 & -70.2 & -43.7 & -0.484 & -0.301 &        &        \\
      1.77 & 145.3 & -63.1 & -36.9 & -0.434 & -0.254 &        &        \\
      1.97 & 145.6 & -64.3 & -37.8 & -0.442 & -0.260 &        &        \\
      2.07 & 145.8 & -67.5 & -40.9 & -0.463 & -0.280 &        &        \\
      2.12 & 145.9 & -64.4 & -37.8 & -0.442 & -0.259 &        &        \\
      2.15 & 146.0 & -67.0 & -40.3 & -0.459 & -0.276 &        &        \\
      2.27 & 145.9 & -62.3 & -35.7 & -0.427 & -0.245 &        &        \\
      3.00 & 141.2 & -61.0 & -35.3 & -0.432 & -0.250 & -0.582 & -0.203 \\
      3.60 & 134.8 & -58.1 & -33.6 & -0.431 & -0.249 & -0.591 & -0.208 \\
      4.27 & 124.0 & -53.3 & -30.9 & -0.429 & -0.249 &        &        \\
  \end{tabular}
  \end{ruledtabular}
  \end{center}
\end{table*}%
for the temperature interval $T=1.0$--4.27~K at~SVP. The corresponding
$^4$He densities are interpolated with the help of data
from~Ref.~\cite{donn98}.  Although transition~II was not observed at gas
densities higher than 32~g/l~\cite{tori99}, the calculated resonance
shifts are also given here for this line, for the sake of comparison.
Note that there are no experimental data about the line shift and
broadening in liquid helium. The density effects have been studied
experimentally only in a~gaseous helium target at pressures
0.2--8.0~bars, temperatures 5.8--6.3~K and target densities ranging from
about 1.4 to~127~g/l \cite{tori99,hori06} where the linear dependence on
the target density has been confirmed within the experimental accuracy.
The recent experimental results of Ref.~\cite{hori06} read
$\Delta{}E_0/\varrho=-0.63\pm{}0.03$~GHz$\cdot$g/l for transition~I and
$\Delta{}E_0/\varrho=-0.21\pm{}0.02$~GHz$\cdot$g/l for transition~II.
Table~\ref{table:shift_sea} shows that the values of the reduced shift
$|\Delta{}E_0|/\varrho$ in liquid $^4$He, calculated using the
phenomenological $\rho_\mathrm{exp}(r)$, differ from both the
experimental data and the semiclassical calculations for gaseous $^4$He.

\begin{figure}[htb]
  \centering
  \includegraphics[width=8cm]{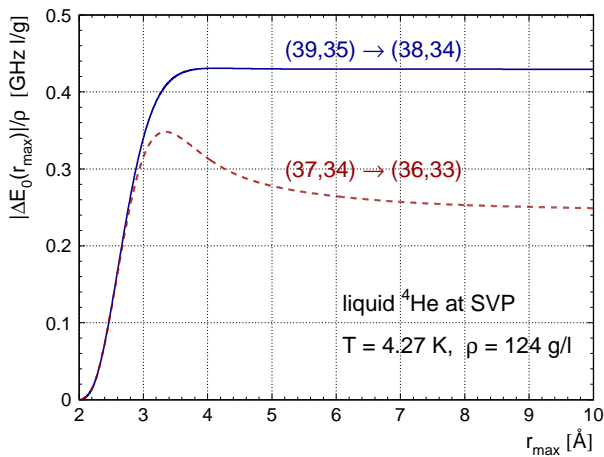}
  \caption{(Color online) The reduced resonance redshifts
    $|\Delta{}E_0|/\varrho$ for transition I and~II as functions
    of~$r_\mathrm{max}$ for $g(r,4.27~\mathrm{K})$
    from~Ref.~\cite{sear79}.
  \label{fig:redshift_sea_hn1}}
\end{figure}

The reduced resonance redshifts $|\Delta{}E_0|/\varrho$ as functions of
the upper limit~$r_\mathrm{max}$ of the integral Eq.~(\ref{qst}) are
plotted in~Fig.~\ref{fig:redshift_sea_hn1}.
In the case of line~I, the asymptotic value of~$\Delta{}E_0$ is
already reached at $r_\mathrm{max}\approx{}3.8$~\AA{}. This confirms
that the approximation of binary interactions is a good one in
this case. A~better approximation would be possible when the
three-body interaction potentials are calculated, which is a~much
more complicated task. For line~II, the asymptotic value
of~$\Delta{}E_0$ is achieved only at
$r_\mathrm{max}\approx{}9$~\AA\ since $\Delta{}V(r)$ changes sign
at $r=3.3$~\AA\ and $\Delta{}E_0(r_\mathrm{max})$ is not a
monotonic function. Thus, in this particular case, the binary
interaction approximation is not as good as for line~I.

\section{Conclusions}
\label{sec:concl}

The present work was motivated by the attempts of the ASACUSA
collaboration at CERN for the high-accuracy spectroscopy measurements of
the antiprotonic helium atoms in liquid helium media, where the
broadening and shift of the spectral lines due to the collective
many-body effects had not been investigated.

The resonance line shift of the antiprotonic helium atom located in
liquid $^4$He is the sum of the contribution $\Delta{}E_0$ from the
pairwise $\bar{p}\mathrm{He}^{+}$-$^4$He interaction and the
contribution $\Delta{}E_1$ due to the collective dynamics of the liquid.
The shift $\Delta{}E_0$ gives a correction on the order of
$|\Delta{}E_0|/E_0\sim{}10^{-4}$ to the resonance wavelength~$\lambda_0$
of an isolated $\bar{p}\mathrm{He}^{+}$ atom. The correction due to the
collective dynamics is much smaller, $|\Delta{}E_1|/E_0\sim{}10^{-6}$
but still may be of importance for high-precision measurements.

The calculated values of $\Delta{}E_0/\varrho$ in
Table~\ref{table:shift_sea} exhibit appreciable variations (9\% for
line~I) with changing temperature, for $T<2.27$~K and almost constant
density. Thus, this phenomenon is apparent only in superfluid~$^4$He. At
higher temperatures, where $^4$He density at SVP significantly
decreases, $\Delta{}E_0/\varrho$ is practically constant. A~similar
effect was observed in the experiments using gaseous helium
targets~\cite{tori99,hori06}.

The reduced line width $\Gamma_1/\varrho$ that comes from the collective
motion in liquid~$^4$He ranges from about $10^{-4}$ to
$10^{-3}$~GHz$\cdot$l/g. Thus, the corresponding contribution to the
line broadening is much lower than the collisional broadening in
a~gaseous helium target as reported in Ref.~\cite{baka00}. As already
pointed out, our method for the evaluation of the density shift in the
quasistatic limit using phenomenological data about the helium target
density cannot be applied in calculating the collisional line
broadening, and we only can estimate the broadening in a~gaseous helium
target as an upper limit for the broadening in liquid helium.

The accuracy of the calculated quasistatic line shifts and broadening
could be improved if the potentials of the $\bar{p}\mathrm{He}^{+}$
interaction with two helium atoms were available. However, the
calculation of such potentials is much more complicated than in the case
of pairwise potential.

Note that one of the methods of studying liquid-helium structure is the
observation of foreign atoms and ions implanted in a liquid $^4$He
target. Unfortunately, the presence of such an atom strongly affects its
helium vicinity. A~large cavity is created around the foreign atom, due
to the Pauli repulsion between the electrons in this atom and the
surrounding helium atoms (see e.g.,~Ref.~\cite{mate11}). The
antiprotonic helium atom, in contrast, having only one electron in the
1s~state, is a good candidate for such studies, as is indicated by the
similarity of the densities $\rho_\mathrm{exp}(r)\equiv{}g(r,T)$ and
$\rho_q(r)$.

\begin{acknowledgments}
  We would like to thank M.~Hori and A.~S\'ot\'er for the valuable
  discussions. This work has been performed under the framework of
  collaboration between the Bulgarian Academy of Sciences and Polish
  Academy of Sciences.
\end{acknowledgments}


\bibliography{ab}

\end{document}